\documentclass[aps,preprint,groupedaddress,floatfix]{revtex4}
\usepackage{graphicx}
\begin{document}

\title{$a_0(980)$ physics in semileptonic $D^0$ and $D^+$ decays }
\author {
N.N. Achasov$^{\,a}$ \email{achasov@math.nsc.ru} and A.V.
Kiselev$^{\,a,b}$ \email{kiselev@math.nsc.ru}}

\affiliation{ $^a$Laboratory of Theoretical Physics,
 Sobolev Institute for Mathematics, 630090, Novosibirsk, Russia\\
$^b$Novosibirsk State University, 630090, Novosibirsk, Russia}

\date{\today}

\begin{abstract}

The decays $D^0\to d\bar u\,e^+\nu\to a^-_0(980)\,
e^+\nu\to\pi^-\eta\, e^+\nu$ and $D^+\to d\bar d\,e^+\nu\to
a^0_0(980)\, e^+\nu\to\pi^0\eta\, e^+\nu$ (and the charge
conjugated ones) are the direct probe of the constituent two-quark
components in the $a^\pm_0(980)$ and $a^0_0(980)$ wave functions.
The recent BESIII experiment is the first step in the experimental
study of these decays. We suggest adequate formulas for the data
analysis and present a variant of $\eta\pi$ invariant mass
distribution when $a_0(980)$ has no constituent two-quark
component at all.

\end{abstract}

\maketitle

\section{Introduction}

The $a_0(980)$ and $f_0(980)$ mesons are well-established parts of
the assumed light scalar meson nonet \cite{pdg-2018}. From the
beginning, the $a_0(980)$ and $f_0(980)$ mesons became one of the
central problems of nonperturbative QCD, as they are important for
understanding the way chiral symmetry is realized in the
low-energy region and, consequently, for understanding
confinement. Many experimental and theoretical papers have been
devoted to this subject.

There is much evidence that supports the four-quark model of light
scalar mesons \cite{jaffe,weinberg}.

The suppression of the $a^0_0(980)$ and $f_0(980)$ resonances in
the $\gamma\gamma\to\eta\pi^0$ and $\gamma\gamma\to\pi\pi$
reactions, respectively, was predicted in the four-quark model in
1982 \cite{fourQuarkGG}, $\Gamma_{a^0_0\gamma\gamma}\approx
\Gamma_{f_0\gamma\gamma}\approx 0.27$ keV, and confirmed by
experiment \cite{pdg-2018}. The high quality Belle data
\cite{uehara2,uehara} allowed one to elucidate the mechanisms of
the $\sigma(600)$, $f_0(980)$, and $a^0_0(980)$ resonance
production in $\gamma\gamma$ collisions \cite{AS88,annsgnGamGam}.
Light scalar mesons are produced in $\gamma\gamma$ collisions
mainly via rescatterings, that is, via the four-quark transitions.
As for $a_2(1320)$ and $f_2(1270)$ (the well-known two-quark
states), they are produced mainly via the two-quark transitions
(direct couplings with $\gamma\gamma$).

The argument in favor of the four-quark nature of $a_0(980)$ and
$f_0(980)$ is the fact that the $\phi(1020)\to a^0_0\gamma$ and
$\phi(1020)\to f_0\gamma$ decays go through the kaon loop:
$\phi\to K^+K^-\to a^0_0\gamma$, $\phi\to K^+K^-\to f_0\gamma$,
i.e., via the four-quark transition
\cite{achasov-89,achasov-97,a0f0,our_a0,achasov-03}. The kaon-loop
model was suggested in Ref. \cite{achasov-89} and confirmed by
experiment ten years later \cite{SNDa0,f0exp,kloea0}.

It was shown in Ref. \cite{achasov-03} that the production of
$a^0_0(980)$ and $f_0(980)$ in $\phi\to a^0_0\gamma\to
\eta\pi^0\gamma$ and $\phi\to f_0\gamma\to\pi^0\pi^0\gamma$ decays
is caused by the four-quark transitions, resulting in strong
restrictions on the large-$N_C$ expansions of the decay
amplitudes. The analysis showed that these constraints give new
evidence in favor of the four-quark nature of the $a_0(980)$ and
$f_0(980)$ mesons.

In Refs. \cite{agsh,ak-07} it was shown that the description of
the $\phi\to K^+K^-\to\gamma a^0_0(980)/f_0(980)$ decays requires
virtual momenta of $K (\bar K)$ greater than $2$ GeV, while in the
case of loose molecules with a binding energy about 20 MeV, they
would have to be about 100 MeV. Besides, it should be noted that
the production of scalar mesons in the pion-nucleon collisions
with large momentum transfers also points to their compactness
\cite{ADS98}.

It was also shown in Refs. \cite{annshgn-94,annshgn-07} that the
linear $S_L(2)\times S_R(2)$ $\sigma$ model \cite{gellman}
reflects all of the main features of low-energy $\pi\pi\to\pi\pi$
and $\gamma\gamma\to\pi\pi$ reactions up to energy 0.8 GeV and
agrees with the four-quark nature of the $\sigma$ meson. This
allowed for the development of a phenomenological model with the
right analytical properties in the complex $s$ plane that took
into account the linear $\sigma$ model, the $\sigma(600)-f_0(980)$
mixing, and the background \cite{our_f0_2011}. This background has
a left cut inspired by crossing symmetry, and the resulting
amplitude agrees with results obtained using the chiral expansion,
dispersion relations, and the Roy equation \cite{sigmaPole}, as
well as with the four-quark nature of the $\sigma(600)$ and
$f_0(980)$ mesons. This model well describes the experimental data
on $\pi\pi\to\pi\pi$ scattering up to $1.2$ GeV.

Moreover, the suppression of $J/\psi\to \gamma f_0(980), \rho
a_0(980), \omega f_0(980)$ decays in the presence of intense
$J/\psi\to \gamma f_2(1270), \gamma f'_2(1525), \rho a_2(1320),
\omega f_2(1270)$ decays is at variance with the $P$-wave
two-quark structure of $a_0(980)$ and $f_0(980)$ resonances
\cite{achasov-1998}.

It is shown in Ref. \cite{correlation} that the recent data on the
$K^0_S K^+$ correlation in Pb-Pb interactions Ref.
\cite{alice-2017} agree with the data on the
$\gamma\gamma\to\eta\pi^0$ and $\phi\to\eta\pi^0\gamma$ reactions
and support the four-quark model of the $a_0(980)$ meson. It is
shown that the data do not contradict the validity of the Gaussian
assumption.

In Refs. \cite{dsdecay,dsdecayConf} the program of studying light
scalars in semileptonic $D$ and $B$ decays was suggested. We
studied production of scalars $\sigma(600)$ and $f_0(980)$ in the
$D_s^+\to\pi^+\pi^-\, e^+\nu$ decays, the conclusion was that the
percentage of the two-quark components in $\sigma(600)$ and
$f_0(980)$ is small. This is the direct evidence in favor of the
exotic nature of these particles. Unfortunately, at the moment the
statistics is rather poor, and thus new high-statistics data are
highly desirable.

It was noted in Refs. \cite{dsdecay,dsdecayConf} that no less
interesting is the study of semileptonic decays of $D^0$ and $D^+$
mesons -- $D^+\to d\bar d\, e^+\nu\to
[\sigma(600)+f_0(980)]e^+\nu\to \pi^+\pi^-e^+\nu$, $D^0\to d\bar
u\, e^+\nu\to a_0^-e^+\nu\to\pi^-\eta e^+\nu$, and $D^+\to d\bar
d\, e^+\nu\to a_0^0 e^+\nu\to\pi^0\eta e^+\nu$ (or the
charged-conjugated ones) which had not been investigated. It is
very tempting to study light scalar mesons in semileptonic decays
of $B$ mesons \cite{dsdecayConf}: $B^0\to d\bar u\, e^+\nu\to
a_0^-e^+\nu\to\pi^-\eta e^+\nu$, $B^+\to u\bar u\, e^+\nu\to a_0^0
e^+\nu\to\pi^0\eta e^+\nu$, $B^+\to u\bar u\, e^+\nu\to
[\sigma(600)+f_0(980)]e^+\nu\to \pi^+\pi^-e^+\nu$.

Recently BES Collaboration measured the decays $D^0\to d\bar u\,
e^+\nu\to a_0^-e^+\nu\to\pi^-\eta e^+\nu$ and $D^+\to d\bar d\,
e^+\nu\to a_0^0 e^+\nu\to\pi^0\eta e^+\nu$ for the first time
\cite{besIII}. In this paper we discuss the Ref. \cite{dsdecay}
program in light of these measurements taking into account the
contribution of the $a_0'$ meson with mass about $1400$ MeV.

A variant when $a_0(980)$ has no constituent two-quark component
at all is presented. That is, $a_0^-(980)$ is produced as a result
of mixing $a_0^{\prime-}\to a_0^-(980)$, $D^0\to d\bar u\,
e^+\nu\to a_0^{\prime -}e^+\nu \to a_0^-e^+\nu\to \pi^-\eta
e^+\nu$, and correspondingly for the $D^+$ decay.

This variant describes the set of experimental data considered in
Ref. \cite{correlation}. Moreover, in comparison with that paper
we take into account high-statistical KLOE data on the
$\phi\to\eta\pi^0\gamma$ decay of Ref. \cite{kloea02009} (instead
of Ref. \cite{kloea0}). To describe this precise data we change
parametrization of the $K\bar K$ scattering background phase,
which changes the module of the $\phi\to
K^+K^-\to(a^0_0+a'^0_0)\gamma\to\eta\pi^0\gamma$ amplitude below
the $K\bar K$ threshold. We also take into account this phase in
the $K^0_S K^+$ correlation and introduce the $m_{a_0^+}$ --
$m_{a_0^0}$ mass difference.

\section{$D$ decays involving scalars and pseudoscalars}

The amplitude of the $D^0\to S (\mbox{scalar)}\, e^+\nu$ decay is
of similar form to the $D^+_s$ decay \cite{dsdecay}
\begin{eqnarray}
 & M[D^0(p)\to S(p_1)W^+(q)\to S(p_1)\, e^+\nu]=\frac{G_F}{\sqrt{2}}V_{cd}A_\alpha
 L^\alpha\,,
 \label{amplitudes}
\end{eqnarray}
where $G_F$ is the Fermi constant, $V_{cd}$ is the
Cabibbo-Kobayashi-Maskava matrix element,
\begin{eqnarray}
 & A_\alpha = f^S_+(q^2)(p+p_1)_\alpha +
f^S_-(q^2)(p-p_1)_\alpha\,,\nonumber \\
 & L_\alpha =\bar{\nu}\gamma_\alpha(1+\gamma_5)e\,,\ \ \ \ \ \ \
 q=(p-p_1)\,.
 \label{VAL}
\end{eqnarray}

The influence of the $f^S_-(q^2)$ form factor is negligible
because of the small mass of the positron.

The decay rate into the stable $S$ state is
\begin{eqnarray}
\frac{d\Gamma(D^0\to
S\,e^+\nu)}{dq^2}=\frac{G^2_F|V_{cd}|^2}{24\pi^3}p^3_1(q^2)|f^S_+(q^2)|^2,
\\ p_1(q^2)=
\frac{\sqrt{m^4_{D^0}-2m^2_{D^0}(q^2+m^2_S)+(q^2-m^2_S)^2
}}{2m_{D^0}}\,.
 \label{dGammadq2}
\end{eqnarray}

For the $f^S_+(q^2)$ form factor we use the vector dominance model
\begin{equation}
f^S_+(q^2)= f^S_+(0)\frac{m^2_A}{m^2_A-q^2}=f^S_+(0)f_A(q^2)\,,
 \label{VDM}
\end{equation}
where $A=D_{1}(2420)^\pm$ \cite{pdg-2018}.

\begin{figure}[h]
\begin{center}
\begin{tabular}{ccc}
\includegraphics[width=8cm,height=5cm]{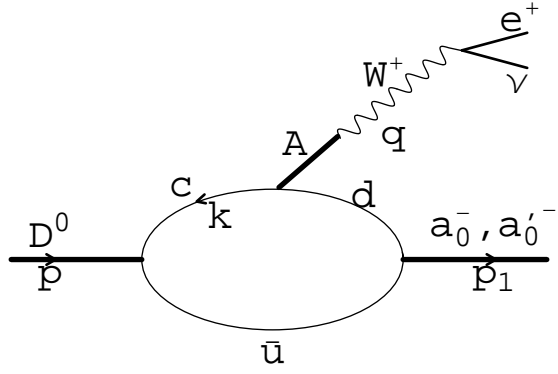}
\end{tabular}
\end{center}
\caption{Model of the $D^0\to (a^-_0,a'^-_0)\, e^+\nu$ decay.}
\label{fig1}
\end{figure}

\begin{figure}[h]
\begin{center}
\begin{tabular}{ccc}
\includegraphics[width=8cm,height=5cm]{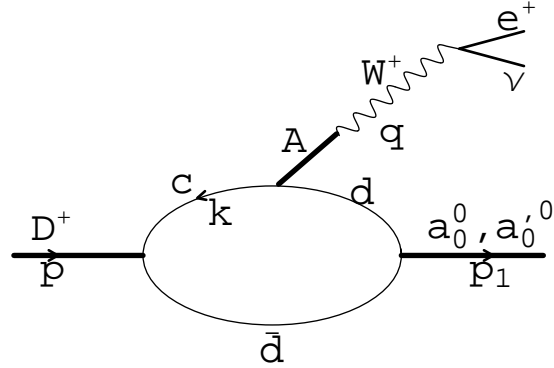}
\end{tabular}
\end{center}
\caption{Model of the $D^+\to (a^0_0,a'^0_0)\, e^+\nu$ decay.}
\label{fig1_1}
\end{figure}

\begin{figure}[h]
\begin{center}
\begin{tabular}{ccc}
\includegraphics[width=8cm,height=8cm]{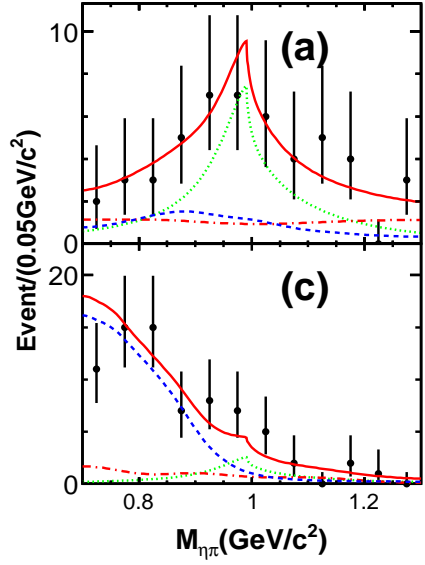}
\end{tabular}
\end{center}
\caption{Experimental data on (a) $D^0\to (a^-_0,a'^-_0)\,
e^+\nu\to\eta\pi^- e^+\nu$ and (c) $D^+\to (a^0_0,a'^0_0)\,
e^+\nu\to\eta\pi^0 e^+\nu$ decays. Direct copy of Figs. 2(a) and
2(c) in Ref. \cite{besIII}. Dotted curves are signals, solid ones
represent total contribution, and the other ones represent
backgrounds.} \label{expData}
\end{figure}

\begin{figure}
\begin{center}
\begin{tabular}{ccc}
\includegraphics[width=8cm]{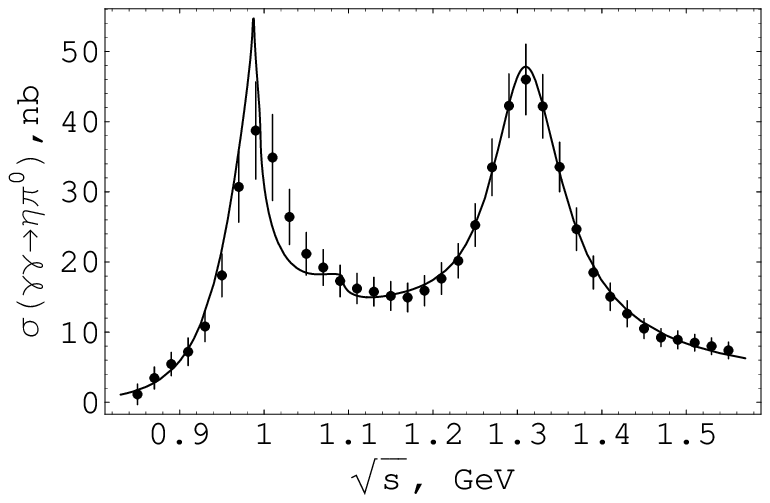}& \includegraphics[width=8cm]{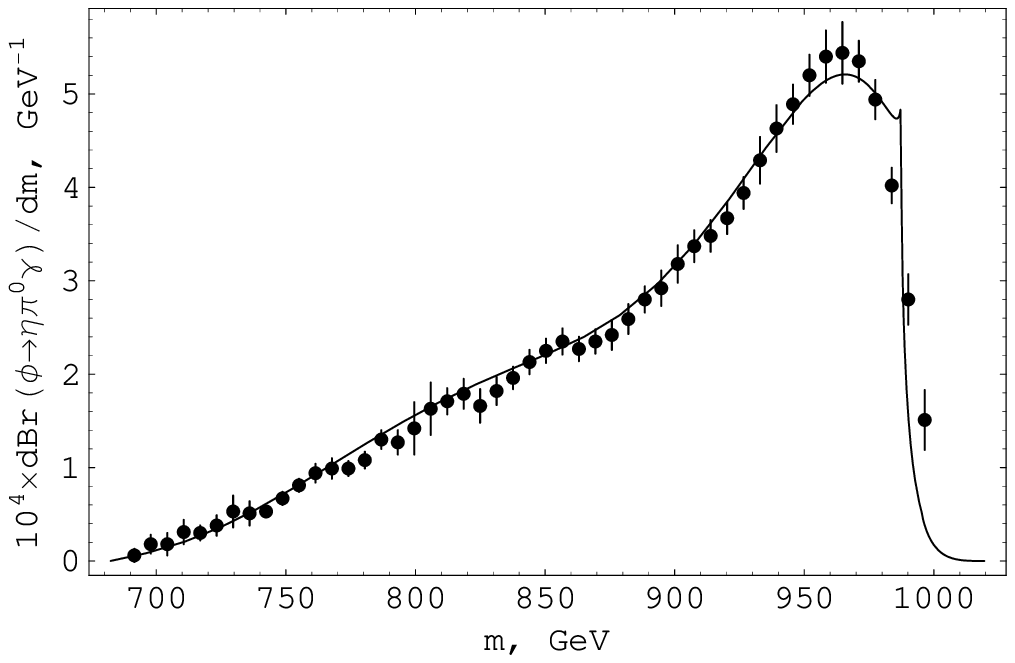}\\ (a)&(b)
\end{tabular}
\end{center}
\caption{Results of our fit (see Tables I and II) on (a) the Belle
data on the $\gamma\gamma\to \eta\pi^0$ cross section
\cite{uehara}, and (b) the KLOE data on the
$\phi\to\eta\pi^0\gamma$ decay \cite{kloea02009}, where $m$ is the
invariant $\eta\pi^0$ mass. } \label{oldData}
\end{figure}

\begin{figure}
\centerline{\includegraphics[width=12cm]{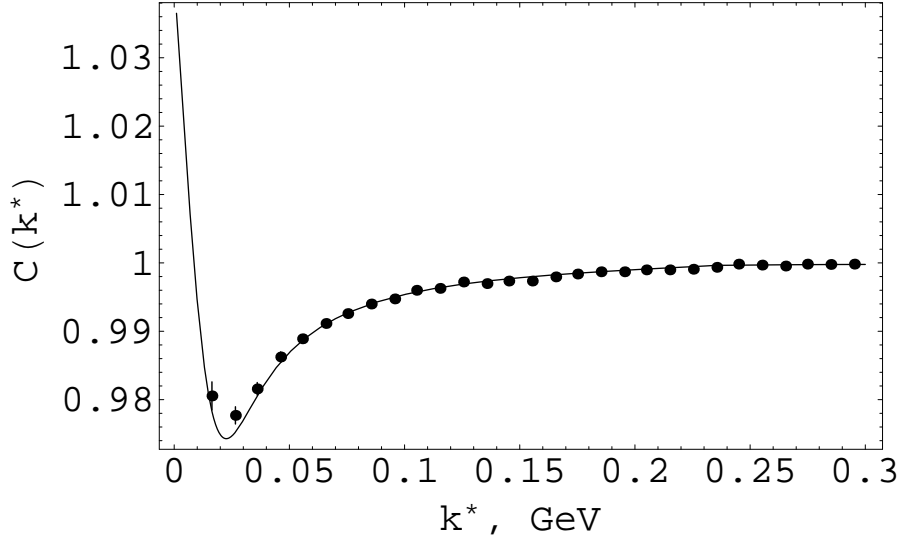}}
\caption {The $K^0_SK^+$ correlation $C(k^*)$; see Ref.
\cite{correlation} and references therein. The solid line
represents our fit, and points are experimental data
\cite{alice-2017}.} \label{corrPlot}\end{figure}

\begin{figure}[h]
\begin{center}
\begin{tabular}{ccc}
\includegraphics[width=8cm]{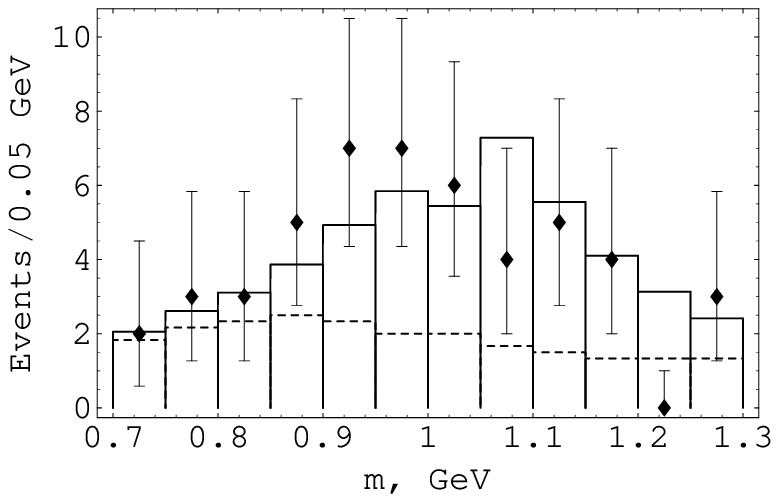}
\end{tabular}
\end{center}
\caption{The data on the $D^0\to (a^-_0,a'^-_0)\,
e^+\nu\to\eta\pi^- e^+\nu$ decay and the fit corresponding to 28.0
events and the signal branching $1.45\times 10^{-4}$. The solid
histogram is the total contribution, and the dashed histogram
represents the sum of backgrounds from Fig. \ref{expData}.}
\label{expDataSuperposed}
\end{figure}

\begin{figure}[h]
\begin{center}
\includegraphics[width=12cm,height=8cm]{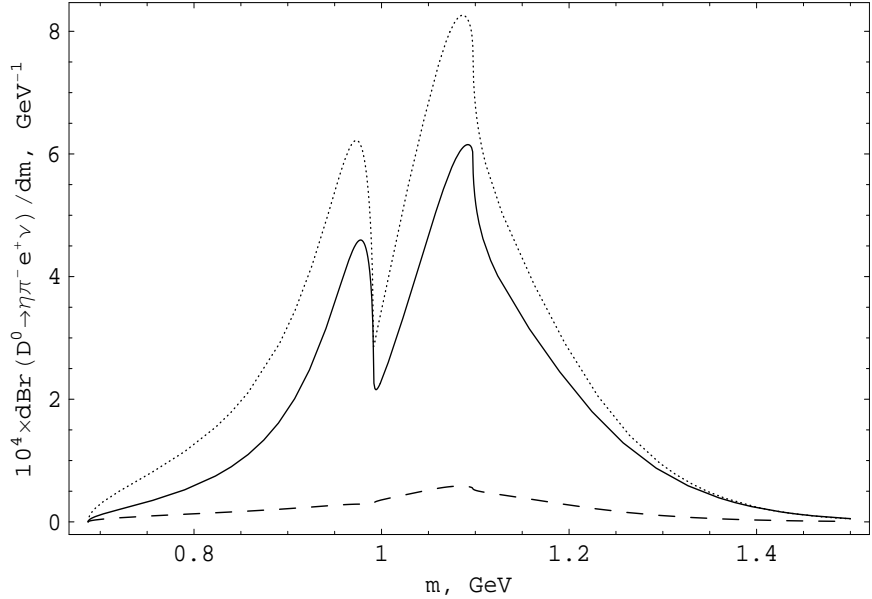}
\end{center}
\caption{The plot of $D^0\to (a^-_0,a'^-_0)\, e^+\nu\to\eta\pi^-
e^+\nu$ spectrum with parameters of our fit (with $g_{d\bar u
a_0^-}=0$). The solid line is the total contribution, the dotted
line is the term $\sim F_{a_0'^-} g_{d\bar u a_0'^-} \Pi_{a_0'^-
a^-_0}(m)g_{a_0\eta\pi}$ contribution, and the dashed line is the
term $\sim F_{a_0'^-} g_{d\bar u a_0'^-}
D_{a^-_0}(m)g_{a_0'\eta\pi}$ contribution; see Eq.
(\ref{d2Gammmadq^2dm}).} \label{dBr1}
\end{figure}

\begin{figure}[h]
\begin{center}
\begin{tabular}{ccc}
\includegraphics[width=8cm]{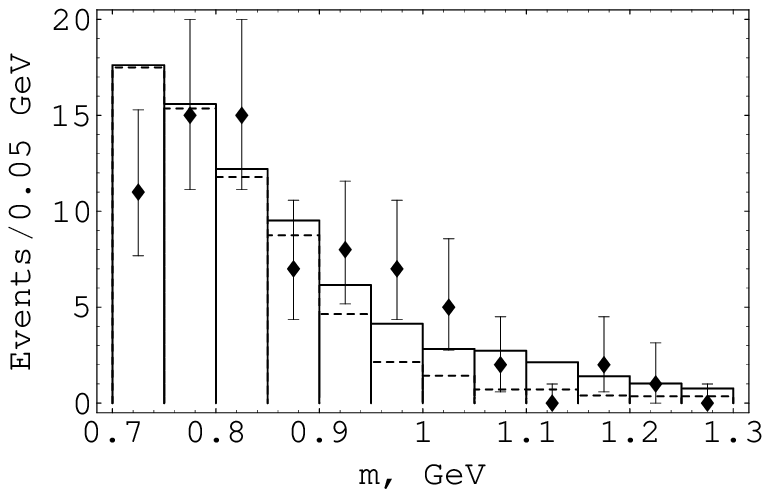}
\end{tabular}
\end{center}
\caption{The data on the $D^+\to (a^0_0,a'^0_0)\,
e^+\nu\to\eta\pi^0 e^+\nu$ decay and the signal corresponding to
fit shown in Fig. \ref{expDataSuperposed}, signal branching is
$1.94\times 10^{-4}$. The solid histogram is the total
contribution, and the dashed histogram represents the sum of
backgrounds from Fig. \ref{expData}.} \label{expDataSuperposedDp}
\end{figure}

\begin{figure}[h]
\begin{center}
\includegraphics[width=12cm,height=8cm]{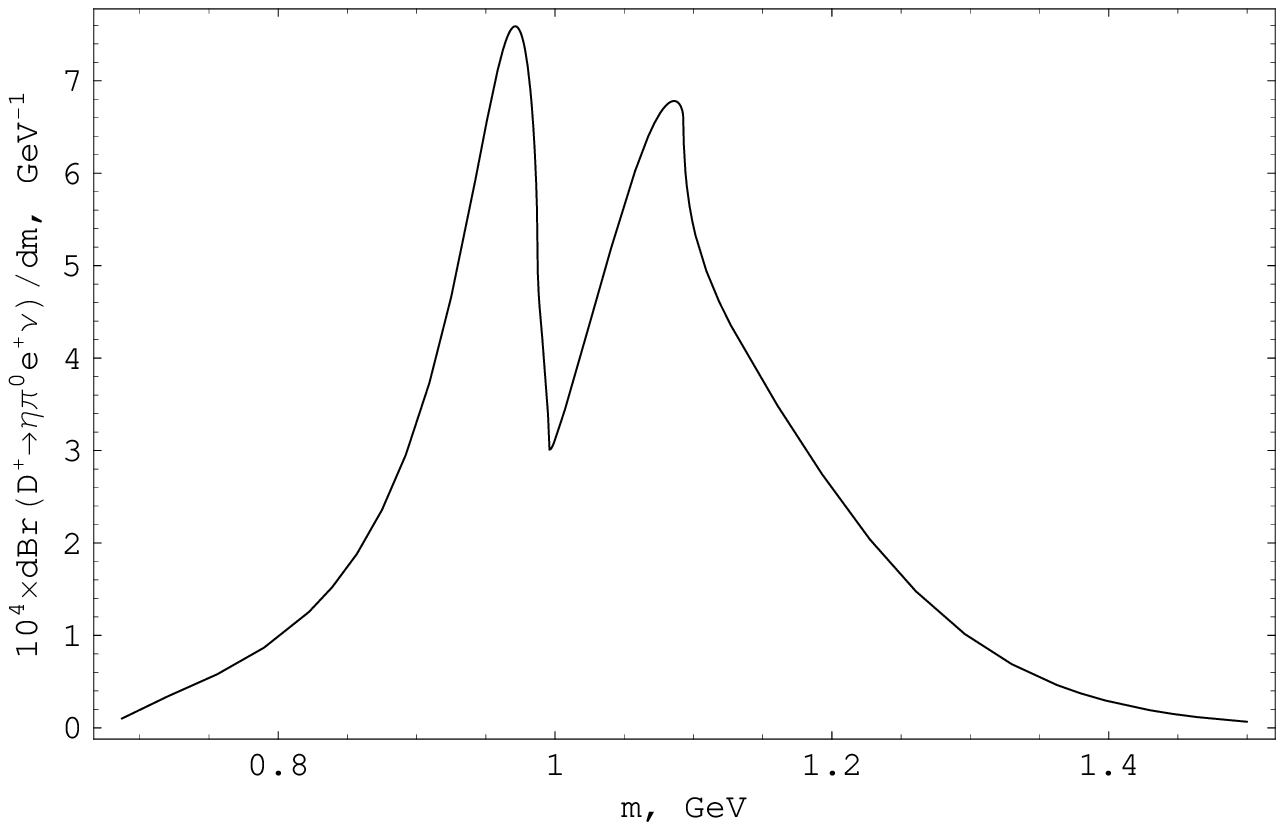}
\end{center}
\caption{The plot of $D^+\to (a^0_0,a'^0_0)\, e^+\nu\to\eta\pi^0
e^+\nu$ spectrum with parameters of our fit.} \label{dBrDp}
\end{figure}

Following Fig. \ref{fig1} we write $f_+^S(0)$ in the form
\begin{equation}
f_+^S(0)= g_{D^0c\bar u}F_Sg_{d\bar uS}\,,
 \label{definitions}
\end{equation}
where $g_{D^0c\bar u}$ is the $D^0\to c\bar u$ coupling constant,
$g_{d\bar uS}$ is the $d\bar u\to S$ coupling constant, and $F_S$
is the loop integral assumed to be constant in the region of
interest.

The amplitude of the $D^0\to d\bar u\, e^+\nu\to
[a_0^-(980)+a'^-_0]\, e^+\nu\to\eta\pi^-\, e^+\nu$ decay is
\begin{eqnarray}
& & M( D^0\to d\bar u\, e^+\nu\, \to\eta\pi^-\, e^+\nu)=
\frac{G_F}{\sqrt{2}}V_{cd}\, L^\alpha\,(p+p_1)_\alpha\,g_{D^0c\bar
u}\,f_A(q^2)\nonumber\\[9pt]
 & &\times \frac{1}{\Delta(m)}\,\Bigl(F_{a^-_0} g_{d\bar u a^-_0}D_{a_0'^-}(m)g_{a_0\eta\pi}+F_{a^-_0} g_{d\bar u a^-_0}
\Pi_{a^-_0 a_0'^-}(m)g_{a_0'\eta\pi}\nonumber\\[9pt] & & \mbox{} +
F_{a_0'^-} g_{d\bar u a_0'^-} \Pi_{a_0'^-
a^-_0}(m)g_{a_0\eta\pi}+F_{a_0'^-} g_{d\bar u a_0'^-}
D_{a^-_0}(m)g_{a_0' \eta\pi}\Bigr )\,,
 \label{D+decayamplitude}
\end{eqnarray}
where $m$ is the invariant mass of the $\eta\pi^-$ system,
$\Delta(m)= D_{a_0'^-}(m)D_{a^-_0}(m)-\Pi_{a_0'^-
a_0^-}(m)\Pi_{a_0^- a_0'^-}(m)$, $D_{a_0^-}(m)$ and
$D_{a_0'^-}(m)$ are the inverted propagators of the $a^-_0$ and
$a_0'^-$ mesons, and $\Pi_{a^-_0 a_0'^-}(m)=\Pi_{a_0'^- a^-_0}(m)$
is the nondiagonal element of the polarization operator, which
mixes the $a_0^-$ and $a_0'^-$ mesons. All the details can be
found in Appendix I.

The double differential rate of the $D^0\to d\bar u\, e^+\nu\to
[a^-_0(980)+a'^-_0]\, e^+\nu\to\eta\pi^-\, e^+\nu$ decay taking
into account the $a'_0$ scalar meson is
\begin{eqnarray}
& & \frac{d^2\Gamma(D^0\to\eta\pi^-\,
e^+\nu)}{dq^2dm}=\frac{G^2_F\,|V_{cd}|^2}{24\,\pi^3}\,g_{D^0c\bar
u}^2\,|f_A(q^2)|^2\,p^3_1(q^2,m)\nonumber\\[9pt] & &
\times\,\frac{1}{8\pi^2}\,m\,\rho_{\eta\pi^-}(m)\,\Bigl|\frac{1}{\Delta(m)}\Bigr|^2\,
\Bigl|F_{a^-_0} g_{d\bar u a^-_0}
D_{a_0'^-}(m)g_{a_0\eta\pi}+F_{a^-_0} g_{d\bar u a^-_0} \Pi_{a^-_0
a_0'^-}(m)g_{a_0'\eta\pi} \nonumber\\[9pt]
 & & \mbox{} + F_{a_0'^-} g_{d\bar u a_0'^-}
\Pi_{a_0'^- a^-_0}(m)g_{a_0\eta\pi}+F_{a_0'^-} g_{d\bar u a_0'^-}
D_{a^-_0}(m)g_{a_0'\eta\pi}\Bigr|^2\,,
 \label{d2Gammmadq^2dm}
\end{eqnarray}
where
$\rho_{\eta\pi^-}(m)=\sqrt{(1-(m_\eta+m_{\pi^-})^2/m^2)(1-(m_\eta-m_{\pi^-})^2/m^2)}$.

The $D^+\to d\bar d\,e^+\nu\to S\, e^+\nu$  and $D^+\to
\eta\pi^0\, e^+\nu$ decays are described in the same way; see Fig.
\ref{fig1_1}. It is enough to substitute in Eqs.
(\ref{amplitudes})-(\ref{d2Gammmadq^2dm}) $D^0\to D^+, d\bar u \to
d \bar d$, $a^-_0\to a^0_0$, $a_0'^- \to a_0'^0$, and
$\pi^-\to\pi^0$. The coupling $g_{d\bar d a'^0_0}=g_{d\bar u
a'^-_0}/\sqrt{2}$.

The key question is the size of the $a_0'$ contribution. In Ref.
\cite{besIII} fits take into account only the $a_0(980)$
contribution, but one can see from Fig. \ref{expData}(a) that the
Ref. \cite{besIII} curve lies below the data in the interval
$m\equiv M_{\eta\pi}=1.1 - 1.3$ GeV (though within large errors).
It may be a manifestation of a sizable $a'_0$ contribution.

In Ref. \cite{correlation} we simultaneously described the data on
the $\gamma\gamma\to\eta\pi^0$ reaction in Ref. \cite{uehara}, the
$\phi\to\eta\pi^0\gamma$ decay \cite{kloea0}, and the recent data
on the $K^0_S K^+$ correlation in Pb-Pb interactions in Ref.
\cite{alice-2017}.

In this article we present for the first time to our knowledge a
variant of data descriptions when $a_0(980)$ has no constituent
two-quark component at all: the $a_0^0(980)$ direct two-quark
transition coupling to the $\gamma\gamma$ channel $g^{(0)}_{a_0
\gamma\gamma}=0$ and $g_{d\bar ua^-_0}=g_{d\bar da^0_0}=0$. The
results are shown in Figs. \ref{oldData} and \ref{corrPlot} and in
Tables I and II.

Fitting the data in Fig. \ref{expData}(a) with obtained parameters
gives the histograms plotted in Figs. \ref{expDataSuperposed} and
\ref{expDataSuperposedDp}. Only normalization is a free parameter
in this fitting. The point on $1.225$ GeV was omitted in fitting,
and the background was extracted from Fig. \ref{expData}(a)
approximately. The optimal integral is 28.0 events in the
experimental region $0.7-1.3$ GeV, and the signal branching
$1.45^{+0.43}_{-0.40}\times 10^{-4}$ -- one can compare it with
Ref. \cite{besIII} result $(1.33^{+0.33}_{-0.29}(stat)\pm
0.09(syst))\times 10^{-4}$. Of course, all this consideration is
very preliminary due to large experimental errors.

The corresponding $dBr(D^0\to d\bar u\,e^+\nu\to (a^-_0,a'^-_0)\,
e^+\nu\to\pi^-\eta\, e^+\nu)/dm$ and $dBr(D^+\to d\bar
d\,e^+\nu\to (a^0_0,a'^0_0)\, e^+\nu\to\pi^0\eta\, e^+\nu)/dm$
curves are shown in Figs. \ref{dBr1} and \ref{dBrDp}. The line
shapes of these curves differ from the signal curve on Figs.
\ref{expData}(a) and \ref{expData}(c).

Some details and parameters of the fit are placed in Appendix II
and Table II therein.

The KLOE data on the $\phi\to\eta\pi^0\gamma$ decay of Ref.
\cite{kloea02009} are so precise that one should take into account
even small effects to describe them. One of the important features
is the background phase of the $K\bar K$ scattering
$\delta^{bg}_{K\bar K}(s)$ for isospin $I=1$, defined in Eqs. (25)
and (27) of Ref. \cite{aks-2015}. Analytical continuation of this
phase under the $K\bar K$ threshold changes the absolute value of
the $\phi\to K^+K^-\to a_0\gamma\to\eta\pi^0\gamma$ amplitude.
Unfortunately, the $K\bar K$ scattering phase is poorly known.

The influence of the analytical continuation of the $K\bar K$
phase is not large near the resonance peak situating near the
$K\bar K$ threshold. In the current work we upgrade the $K\bar K$
scattering phase parametrization:
\begin{equation}
e^{2i\delta^{bg}_{K\bar K}(s)}=\frac{1+iF_{K\bar
K}(s)}{1-iF_{K\bar K}(s)},\hspace{20pt} e^{2i\delta^{bg}_{\bar K^0
K^+}(s)}=\frac{1+iF_{\bar K^0 K^+}(s)}{1-iF_{\bar K^0 K^+}(s)}\,,
\end{equation} where
\begin{equation}
F_{K\bar K}(s)=f_{K\bar
K}\frac{\sqrt{s-4m_{K^+}^2}+\sqrt{s-4m_{K^0}^2}}{2}+g_{K\bar
K}\frac{\sqrt{1-4m_{K^+}^2/s}+\sqrt{1-4m_{K^0}^2/s}}{2}\,,\label{KKphase}\end{equation}

$$F_{\bar K^0 K^+}(s)=f_{K\bar
K}\frac{\sqrt{(s-(m_{K^0}+m_{K^+})^2)(s-(m_{K^0}-m_{K^+})^2)}}{\sqrt{s}}+
$$ \begin{equation} g_{K\bar
K}\frac{\sqrt{(s-(m_{K^0}+m_{K^+})^2)(s-(m_{K^0}-m_{K^+})^2)}}{s}\,.\label{KKphase2}\end{equation}

Compared with parametrization used in \cite{aks-2015} and later,
we add to $F_{K\bar K}(s)$ a term proportional to velocity and
take into account the kaon mass difference. The phase
$\delta^{bg}_{K\bar K}$ is used in the $\gamma\gamma\to\eta\pi^0$
and $\phi\to\eta\pi^0\gamma$ reactions, and $\delta^{bg}_{\bar K^0
K^+}$ is used to study the $K^0_S K^+$ correlation.

We also upgrade Eq. (6) in Ref. \cite{correlation} describing the
amplitude of the $\bar K^0 K^+$ scattering:
\begin {equation} f(k^*)=
\frac{e^{2i\delta^{bg}_{\bar K^0 K^+}(s)}-1}{2i\rho_{K^0
K^+}}+e^{2i\delta^{bg}_{\bar K^0
K^+}(s)}\frac{4}{\sqrt{s}}\sum_{S,S'}\frac{g_{S K^0_S
K^+}G_{SS'}^{-1}g_{S'K^0_S K^+}}{16\pi},
\end{equation}

\noindent where $S,S'=a^+_0,a'^+_0$, the constants $g_{S K^0_S
K^+}=-g_{S K^0_L K^+}=g_{SK^+K^-}$, and $k^*$ is the kaon momentum
in the kaon pair rest frame,
\begin {equation}
k^*=\frac{\sqrt{(s-(m_{K^0_S}-m_{K^+})^2)(s-(m_{K^0_S}+m_{K^+})^2)}}{2\sqrt{s}}.
\end{equation}

In comparison with Eq. (6) of Ref. \cite{correlation} we take the
$\bar K^0 K^+$ scattering phase into account and fix a misprint:
$\frac{2}{\sqrt{s}}$ was written instead of $\frac{4}{\sqrt{s}}$.
The calculations in Ref. \cite{correlation} were done with the
correct formula.

Remember that the $K^0_S K^+$ correlation reads
\cite{alice-2017,ledlub-1982}
\begin {equation}
C(k^*)=1+\frac{\lambda}{2}\bigg(\frac{1}{2}\bigg|\frac{f(k^*)}{R}\bigg|^2+2\frac{Re
f(k^*)}{\sqrt{\pi}R}F_1(2k^*R)-\frac{Im
f(k^*)}{R}F_2(2k^*R)\bigg),\label{corrForm}
\end{equation} where $R$ is the radius parameter from the spherical Gaussian source distribution, $\lambda$ is the correlation
strength, and  \begin{equation}F_1(z)=\frac{e^{-z^2}}{z}\int_0^z
e^{x^2}dx;\hspace{5pt} F_2(z)=\frac{1-e^{-z^2}}{z}.\label{F12}
\end{equation}

\begin{center}
Table I. Properties of the resonances and the description quality.
\begin{tabular}{|c|c|c|c|c|c|c|}\hline

$m_{a^0_0}$, MeV & $988.3$ & $m_{a'_0}$, MeV & $1423.9$ & $R$, fm
& $6.3$ \\ \hline

$g_{a^0_0K^+K^-}$, GeV  & $4.06$ & $g_{a_0'^0 K^+K^-}$, GeV  &
$4.19$ & $\lambda$ & $1$
\\ \hline

$g_{a_0 \eta\pi}$, GeV  & $3.99$ & $g_{a_0' \eta\pi}$, GeV  &
$0.80$ & $\chi^2_{\gamma\gamma}$ / $36$ points & $13.8$ \\ \hline

$g_{a_0 \eta'\pi}$, GeV  & $-4.24$ & $g_{a_0' \eta'\pi}$, GeV  &
$1.27$ & $\chi^2_{sp}$ / $49$ points & $65.5$
\\ \hline

$g^{(0)}_{a^0_0 \gamma\gamma}$ & $0$ & $g^{(0)}_{a_0'^0
\gamma\gamma}$, $10^{-3}\,$GeV$^{-1}$ & $-12.90$ & $\chi^2_{corr}$
/ $29$ points & $28.4$
\\ \hline

$m_{a^+_0}$, MeV & $997.6$ & $C_{a_0 a'_0}$, GeV$^2$ & $-0.163$ &
($\chi^2_{\gamma\gamma}$+$\chi^2_{sp}$+$\chi^2_{corr}$)/n.d.f. &
$107.8/99$
\\ \hline

\end{tabular}
\end{center}

To fit the data we use the $\chi^2$ function with the addition of
terms providing some restrictions, including terms that guarantee
being close to the four-quark model relations; see Appendix 3 in
Ref. \cite{aks-2015} for details. Finally there are 15 effective
free parameters of the fit, including several parameters that are
softly restricted by terms $\sim (P-P_{0})^2$, where $P$ is the
parameter and $P_{0}$ is its notably desired value. So results in
Tables I and II are not obtained by pure $\chi^2$ method, and we
present a possible scenario.

For the data on $\phi\to\eta\pi^0\gamma$ we use a modified
$\chi^2$ function stressing on the resonant region $m>800$ MeV and
with poor weight of the low $m$ region. One can see in Fig.
\ref{oldData}(b) that the description is close to experimental
data for all $m$.

We faced several minima of the resulting function to minimize;
they are rather close to each other. We show the best one.
$\chi^2_{\gamma\gamma}$, $\chi^2_{sp}$, and $\chi^2_{corr}$ shown
in Table I are pure $\chi^2$ values built on
$\gamma\gamma\to\eta\pi^0$ data \cite{uehara}, the data on the
$\phi\to\eta\pi^0\gamma$ decay \cite{kloea02009}, and the $K^0_S
K^+$ correlation data \cite{alice-2017} correspondingly.

Since we use a different model (including different
parametrization of $\delta^{bg}_{K\bar K}$ and $\delta^{bg}_{\bar
K^0 K^+}$) and different data set (newer KLOE data on
$\phi\to\eta\pi^0\gamma$ decay), and, besides, consider the case
when $a_0(980)$ has no constituent two-quark component, the
results shown in Tables I and II differ from the results in Ref.
\cite{correlation}. One can treat this difference as an error
estimation.

\section{Conclusion}

The first measurement of $D^0\to d\bar u\,e^+\nu\to
[a^-_0(980)+a'^-_0]\, e^+\nu\to\pi^-\eta\, e^+\nu$ and $D^+\to
d\bar d\,e^+\nu\to [a^0_0(980)+a_0'^0]\, e^+\nu\to\pi^0\eta\,
e^+\nu$ decays \cite{besIII} is an important step for the
investigation the nature of of light scalar mesons.

The data description with $g^{(0)}_{a^0_0 \gamma\gamma}=0$ is
presented for the first time to our knowledge, and it means that
$a_0(980)$ has no constituent two-quark component. The data are
described well, and the $a_0(980)$ coupling constants agree with
the four-quark model scenario: they obey (or almost obey) the
relations \cite{achasov-89}

\begin{eqnarray} &
g_{a_0\eta\pi^0}=\sqrt{2}
\mbox{sin}(\theta_p+\theta_q)g_{a_0K^+K^-}=(0.85 - 0.98)
g_{a_0K^+K^-} , & \nonumber\\ & g_{a_0\eta'\pi^0}=-\sqrt{2}
\mbox{cos}(\theta_p+\theta_q)g_{a_0K^+K^-}=-(1.13 - 1.02)
g_{a_0K^+K^-} , & \label{FourQrelations}
\end{eqnarray}

\noindent where $g_{a_0\eta\pi^0}=0.85\, g_{a_0K^+K^-}$ and
$g_{a_0\eta'\pi^0}=-1.13\, g_{a_0K^+K^-}$ for $\theta_p=-18^\circ$
and $g_{a_0\eta\pi^0}=0.98\, g_{a_0K^+K^-}$ and
$g_{a_0\eta'\pi^0}=-1.02\, g_{a_0K^+K^-}$ for
$\theta_p=-11^\circ$. The $\theta_q=54.74^\circ$.

The corresponding prediction of $D^0\to d\bar u\,e^+\nu\to
[a^-_0(980)+a'^-_0]\, e^+\nu\to\pi^-\eta\, e^+\nu$ and $D^+\to
d\bar d\,e^+\nu\to [a^0_0(980)+a_0'^0]\, e^+\nu\to\pi^0\eta\,
e^+\nu$ decays is presented and does not contradict the data
\cite{besIII}. An experiment on higher statistics could check this
prediction.

The experiment on $D_s^+\to s\bar s\, e^+\nu\to
[\sigma(600)+f_0(980)+f'_0]\, e^+\nu\to\pi^+\pi^-\, e^+\nu$ with
higher precision than in Ref. \cite{cleo} is also strongly
interesting.

Let us repeat that no less interesting is to probe the light
scalars in semileptonic $D^+\to d\bar d\, e^+\nu\to
[\sigma(600)+f_0(980)+f'_0]e^+\nu\to \pi^+\pi^-e^+\nu$, $B^0\to
d\bar u\, e^+\nu\to [a^-_0(980)+a'^-_0]e^+\nu\to\pi^-\eta e^+\nu$,
$B^+\to u\bar u\, e^+\nu\to [a_0^0(980)+a_0'^0] e^+\nu\to\pi^0\eta
e^+\nu$, and $B^+\to u\bar u\, e^+\nu\to
[\sigma(600)+f_0(980)+f'_0]e^+\nu\to \pi^+\pi^-e^+\nu$ decays
which have not yet been investigated.

The approach of this paper is valid for the $B$ mesons decays
$B^0\to\pi^-\eta e^+\nu$ and $B^+\to \pi^0\eta e^+\nu$. It is
enough to make obvious changes $V_{cd}\to V_{ub}$, $g_{D^0c\bar
u}\to g_{B^0d\bar b}$, and $g_{D^+c\bar d}\to g_{B^+u\bar b}$. In
Eq. (\ref{VDM}) $A=D_{1}(2420)^\pm\to B_1(5721)^+$.

\section{Appendix I: Scalar propagators and polarization operators}
\label{polarizationOp}

The matrix of the inverse propagators is
\begin{equation}G_{SS'}(m)=\left(
\begin{array}{cc}
D_{a'_0}(m)&-\Pi_{a'_0a_0}(m)\\-\Pi_{a'_0a_0}(m)&D_{a_0}(m)\end{array}\right),\end{equation}
\begin{equation}\Pi_{a'_0a_0}(m)=\sum_{a,b} \frac{g_{a'_0 ab}}{g_{a_0
ab}}\Pi^{ab}_{a_0}(m)+C_{a'_0 a_0},\end{equation} \noindent where
$m=\sqrt{s}$, and the constant $C_{a_0'a_0}$ incorporates the
subtraction constant for the transition $a_0(980)\to(0^-0^-)\to
a_0'$ and effectively takes into account the contributions of
multiparticle intermediate states to the $a_0\leftrightarrow a_0'$
transition. The inverse propagator of the scalar meson $S$
\cite{achasov-89,achasov-97,aks-2015,adsh-79} is
\begin{equation} \label{propagator} D_S(m)=m_S^2-m^2+\sum_{ab}[Re
\Pi_S^{ab}(m_S^2)-\Pi_S^{ab}(m^2)],
\end{equation}
\noindent where $\sum_{ab}[Re \Pi_S^{ab}(m_S^2)-
\Pi_S^{ab}(m^2)]=Re\Pi_S(m_S^2)- \Pi_S(m^2)$ takes into account
the finite-width corrections of the resonance which are the
one-loop contributions to the self-energy of the $S$ resonance
from the two-particle intermediate  $ab$ states. We take into
account the intermediate states $\eta\pi^+,K\bar K$, and
$\eta'\pi^+$ in the $a^+_0(980)$ and $a_0'^+$ propagators:
\begin{equation}
\Pi_S=\Pi_S^{\eta\pi^+}+\Pi_S^{K^0_SK^+}+\Pi_S^{K^0_LK^+}+
\Pi_S^{\eta'\pi^+},
\end{equation}

\noindent and $\eta\pi^0,K\bar K$, and $\eta'\pi^0$ in the
$a^0_0(980)$ and $a_0'^0$ propagators.

For pseudoscalar mesons $a,b$ and $m_a\geq m_b,\ m\geq m_+$, one
has \begin{eqnarray} \label{polarisator}
&&\Pi^{ab}_S(m^2)=\frac{g^2_{Sab}}{16\pi}\left[\frac{m_+m_-}{\pi
m^2}\ln \frac{m_b}{m_a}+\right.\nonumber\\
&&\left.+\rho_{ab}\left(i+\frac{1}{\pi}\ln\frac{\sqrt{m^2-m_-^2}-
\sqrt{m^2-m_+^2}}{\sqrt{m^2-m_-^2}+\sqrt{m^2-m_+^2}}\right)\right]
,\end{eqnarray}

\noindent where
$\rho_{ab}(s)=2p_{ab}(s)/\sqrt{s}=\sqrt{(1-m_+^2/s)(1-m_-^2/s)}$,
and $m_\pm=m_a\pm m_b$. Analytical continuation to other energy
regions could be found, for example, in Ref. \cite{correlation}
and references therein.

The constants $g_{Sab}$ are related to the width as
\begin{equation}
\Gamma_S(m)=\sum_{ab} \Gamma(S\to
ab,m)=\sum_{ab}\frac{g_{Sab}^2}{16\pi m}\rho_{ab}(m). \label{GRab}
\end{equation}

\section{Appendix II: Other parameters and details}

For completeness, we show in Table II the background parameters
and the parameters that are not described above. One can find all
of the details in Ref. \cite{aks-2015}.

\begin{center}
Table II. Parameters not mentioned in Table I.

\begin{tabular}{|c|c|c|c|c|c|}\hline

$c_0$ & $-0.34$ & $f_{K\bar K}$, GeV$^{-1}$ & $-2.14$ \\ \hline
$c_1$, GeV$^{-2}$ & $-9.04$ & $g_{K\bar K}$ & $2.37$
\\ \hline

$c_2$, GeV$^{-4}$ & $1.40$ & $f_{\pi\eta'}$, GeV$^{-1}$ & $-0.50$
\\ \hline

$\delta,^{\circ}$ & $-128.3$ &  &
\\ \hline

\end{tabular}
\end{center}

In this paper we take the form factor $G_\omega (s,t)=G_\rho
(s,t)$,
\begin{equation} G_\omega(s,t)=G_\rho
(s,t)=\exp[(t-m^2_\omega)b_\omega (s)]\,, \label{Gfactor}
\end{equation} differently from Refs.
\cite{annsgnGamGam,aks-2015}. We take \begin{equation}
b_\omega(s)=b^0_\omega+\alpha'_\omega \ln[1+(s/s_0)]
\end{equation}

\noindent and obtain $b^0_\omega=2.3\times 10^{-3}$ GeV$^{-2}$,
and $s_0=1.005$ GeV$^2$. $\alpha'_\omega=0.8$ GeV$^{-2}$ is the
same. Form factors for the $K^*$ exchange are modified the same
way. Besides, we obtain $r_{a_2}=1.2$ GeV$^{-1}$ instead of
$r_{a_2}=1.9$ GeV$^{-1}$ in Refs. \cite{annsgnGamGam,aks-2015}.

The $\pi\eta$ scattering length agrees with the estimates based on
current algebra and chiral perturbation theory, according to which
$a^1_0\approx 0.005-0.01$ (in units of $m^{-1}_\pi$); see Ref.
\cite{annsgnGamGam}.

\section{Acknowledgements}

The work was supported by the program of fundamental scientific
researches of the SB RAS No. II.15.1., Project No. 0314-2016-0021.
The present work is partially supported by the Russian Foundation
for Basic Research Grant No. 16-02-00065.

\end{document}